\documentclass[aps,pra,superscriptaddress,twocolumn,showpacs,floatfix]{revtex4}
\usepackage{amsmath,amssymb,amsfonts,bbm,graphicx,color, setspace, float, textcomp,hyperref,cancel}
 \newcommand{\ket}[1]{\vert #1 \rangle}
 
\def\i{{rm i}}

 \def\e{\mbox{e}}

\def\omegap{{\omega_{p}}}

\def\e{\text{e}}
\def\i{\text{i}}

\begin{document}
\title{High-order dispersion effects in two-photon interference}
\author{Zeudi Mazzotta}\email{zeudi.mazzotta@unimi.it}
\affiliation{Quantum Technology Lab, Dipartimento di Fisica, Universit\`a degli Studi di Milano, I-20133 Milano, Italy}
\affiliation{Istituto Nazionale di Fisica Nucleare, Sezione di 
Milano, Via Celoria 16, I-20133 Milan, Italy}
\author{Simone Cialdi}\email{simone.cialdi@unimi.it}
\affiliation{Quantum Technology Lab, Dipartimento di Fisica, Universit\`a degli Studi di Milano, I-20133 Milano, Italy}
\affiliation{Istituto Nazionale di Fisica Nucleare, Sezione di 
Milano, Via Celoria 16, I-20133 Milan, Italy}
\author{Daniele Cipriani}\email{daniele.cipriani@mi.infn.it}
\affiliation{Quantum Technology Lab, Dipartimento di Fisica, Universit\`a degli Studi di Milano, I-20133 Milano, Italy}
\author{Stefano Olivares}\email{stefano.olivares@fisica.unimi.it}
\affiliation{Quantum Technology Lab, 
Dipartimento di Fisica, Universit\`a degli Studi di Milano, I-20133 Milano, Italy}
\affiliation{Istituto Nazionale di Fisica Nucleare, Sezione di 
Milano, Via Celoria 16, I-20133 Milan, Italy}
\author{Matteo G. A. Paris}\email{matteo.paris@fisica.unimi.it}
\affiliation{Quantum Technology Lab, Dipartimento di Fisica, 
Universit\`a degli Studi di Milano, I-20133 Milano, Italy}
\affiliation{Istituto Nazionale di Fisica Nucleare, Sezione di 
Milano, Via Celoria 16, I-20133 Milan, Italy}
\date{\today}
\begin{abstract}
Two-photon interference and Hong-Ou-Mandel (HOM) effect are relevant
tools for quantum metrology and quantum information processing.  In
optical coherence tomography, HOM effect is exploited to achieve
high-resolution measurements with the width of the HOM dip being the
main parameter. On the other hand, applications like dense coding require
high-visibility performances. Here we address high-order dispersion
effects in two-photon interference and study, theoretically and
experimentally, the dependence of the visibility and the width of 
the HOM dip on {\em both} the pump spectrum
and the downconverted photon spectrum. In particular, a spatial light
modulator is exploited to experimentally introduce and manipulate a
custom phase function to simulate the high-order dispersion effects.
\end{abstract}
\maketitle
\section{Introduction}
Two-photon interference (TPI) and in particular the the Hong-Ou-Mandel
(HOM) \cite{hom87} effect are relevant tools in photonic quantum technology.  The HOM
dip plays a central role in dense and superdense coding, since it allows
to identify two of the four Bell states \cite{hw94,slb95}. Here the
crucial parameter is the visibility of the HOM dip: the higher the
visibility, the higher the mutual information \cite{tim1,tim2}.  TPI
have found applications also in quantum metrology, where frequency
entangled states of photons produced by downconversion 
(PDC) represent a resource for clock synchronization \cite{gio01}
and optical coherence
tomography \cite{hu91}. However, the need to improve the resolution
performances while using low-coherence interference leads to a
halt: if one increases resolution widening the spectrum, the effect of
dispersive media all along the path gets larger, thus preventing any
further improvement \cite{eng98}. Nevertheless, since the width of
the HOM dip is not affected by dispersive media \cite{tim5}, TPI represents a
promising candidate to face the resolution puzzle \cite{tim3,tim4}: the
so-called Quantum Optical Coherence Tomography \cite{abo2002,nas2003}
represents an interferometric technique for axial imaging
offering several advantages over conventional methods. 
\par
This feature of the HOM effect, i.e. the independence of  the dip width on dispersion,
however, holds only if the  pump laser is nearly monochromatic. When a broadband
source is used to produce frequency entangled photons, the HOM dip becomes sensitive to
{even} order dispersions 
\cite{res2009}, and both width and visibility are degraded.
\par
In view of its relevance for quantum technology \cite{tim5,mat96}, theoretical
\cite{tim6,per99,erd2000} and experimental \cite{tim3,tim4,tim7,tim8}
studies have been performed to characterize TPI and the HOM dip. 
However, a full analysis of both the visibility and the width of the HOM dip as a function 
of the pump spectrum and of the PDC spectrum is still missing.
\par
The dispersive effects on the {propagating} photons can be described by
a suitable phase function: this is the main topic of the present paper.
In particular, we develop an experimental apparatus to introduce and
manipulate a custom phase function with a spatial light modulator, as
well as to work with different pump and PDC spectrum. As a matter of
fact, we can
simulate high-order dispersive effects of different transmission
channels {by} exploiting the same setup. Indeed, we have performed a
complete theoretical and experimental study of how the HOM effect is
affected by the presence of second and third order
dispersions, together with a direct comparison between 
theoretical and
experimental results. The same setup can be used to 
compensate this kind of effect, paving the way 
for optimization in quantum technological applications 
involving PDC photons and optical fibers.
\par
The paper is structured as follows. In Section \ref{s:tpi} we review
the description of frequency entangled photons obtained by PDC and
provide the basic principles of two-photon interference in 
dispersive media. In Section \ref{s:expa} we describe in some details
our experimental apparatus, whereas Section \ref{s:expr} is devoted to
the experimental results. We discuss our findings in Section \ref{s:discuss} and
close the paper with some concluding remarks in Section \ref{s:out}.

\section{Two-photon interference}\label{s:tpi}
The HOM effect occurs in TPI when two frequency-entangled
photons generated by PDC interfere at a beam splitter (BS).
In order to describe the TPI process after the propagation in 
dispersive media, we need to introduce the two-photon
state produced via PDC and traveling from the source to the BS. 
In the presence of a broadband source laser
the quantum state of the two photons, i.e. the signal ``$s$'' and the idler ``$i$''
photons, which interfere at the BS, is given by (see, e.g., \cite{smi2013}):
\begin{align}
\ket{\Phi} = \int\!\!\!\int\! & d\omegap\,d\omega\, A(\omegap)\, f(\omega)\, \nonumber \\
& \times \e^{\i \phi_{s}(\omegap,\omega)+\i \phi_{i}(\omegap,-\omega)}
\ket{\omegap,\omega}_s
\ket{\omegap,-\omega}_i,\label{si}
\end{align}
where $\omegap$ and $\omega$ are the frequency shifts
from the central pump and the central PDC frequencies, respectively,
whereas $A(\omegap)$ and $f(\omega)$ are the pump and the
PDC spectral amplitudes, respectively.
In Eq.~(\ref{si}) we introduced the two phase functions
$\phi_{s}(\omegap,\omega) \equiv \phi_{s}(\omega_s)$ and
$\phi_{i}(\omegap,\omega) \equiv \phi_{i}(\omega_i)$ 
denoting the phase functions induced by the propagation of the signal and
the idler photon with frequencies
$\omega_{s} = \left(\frac{\omegap}{2}+\omega\right)$ and 
$\omega_{i} = \left(\frac{\omegap}{2}-\omega\right)$, respectively.
Since, in general,  $\omegap \ll \omega$, the effects of
the broadband pump on $f(\omega)$ are negligible.
\par
The phase functions are responsible for high-order effects on the
propagation of state. In order to enlighten the different contributions,
we expand $\phi_{x}(\omega_x)$ up to the third order in $\omega_x$, $x=s,i$,
namely:
\begin{align}
 \phi_{x}(\omega_x) &\approx
 \sum_{k=0}^{3} \frac{1}{k!}\, \frac{\partial^n }{\partial\omega_x^n}\phi_x(0)\, \omega_x^n\\
 &\approx \beta^{(0)}_{x}+ \beta^{(1)}_{x}\omega_{x}  +
 \frac{\beta^{(2)}_{x}}{2!}\omega_{x}^2 +
 \frac{\beta^{(3)}_{x}}{3!}\omega_{x}^3,
 \end{align}
where $\beta^{(k)}_{x} = \partial^n_{\omega_x} \phi_x(0)$.
Each coefficient $\beta^{(k)}_{x}$ of the expansion plays
a well defined role in the photon propagation (in our analysis we do not consider the constant phase
$\beta^{(0)}_{x}$ since, as we will see, it will not contribute to the quantities of interest considered
throughout the rest of the paper). The first order coefficient $\beta^{(1)}_{x}$ represents the ``time''
spent by the photons to reach the BS and is related to the group velocity.
The second order coefficient $\beta^{(2)}_{x}$ refers to the group velocity dispersion
experienced by both photons and it is due to the presence of dispersive media. Finally,
the third order dispersion coefficient $\beta^{(3)}_{x}$, though it has 
usually a small impact in common materials, may have a huge effect when using
very long optical fibers.
\par
In each path, then, the two photons can experience a different 
dispersion, which may largely affect the TPI process and the 
occurrence of the HOM effect. More in details, if we introduce the ``delay time''
$\tau =\beta^{(1)}_{s} - \beta^{(1)}_{i}$, the probability of having 
coincident counts after the BS as a function $\tau$ reads:
 \begin{align}
 P(\tau) = \int\!\!\!\int\!  d\omegap\,  & d\omega\,
 |A(\omegap)|^2 \,
 \bigg| T f(\omega) \e^{\i [\phi_s(\omegap,\omega)+\phi_i(\omegap,-\omega)]}
 \nonumber\\
 &-R f(-\omega) \e^{\i [\phi_s(\omegap,-\omega)+\phi_i(\omegap,\omega)]} \bigg|^2,
 \label{eq:Ptau}
 \end{align}
 where $T$ and $R$ are the transmission and reflection 
 coefficients of the beam splitter.

\par
In the following we will show how we can experimentally manipulate and control
the different contributions described above and, in particular, we will investigate their
effects on the HOM dip and visibility.
\section{Experimental apparatus}\label{s:expa}
\begin{figure}[tb!]
\includegraphics[width=0.99\columnwidth]{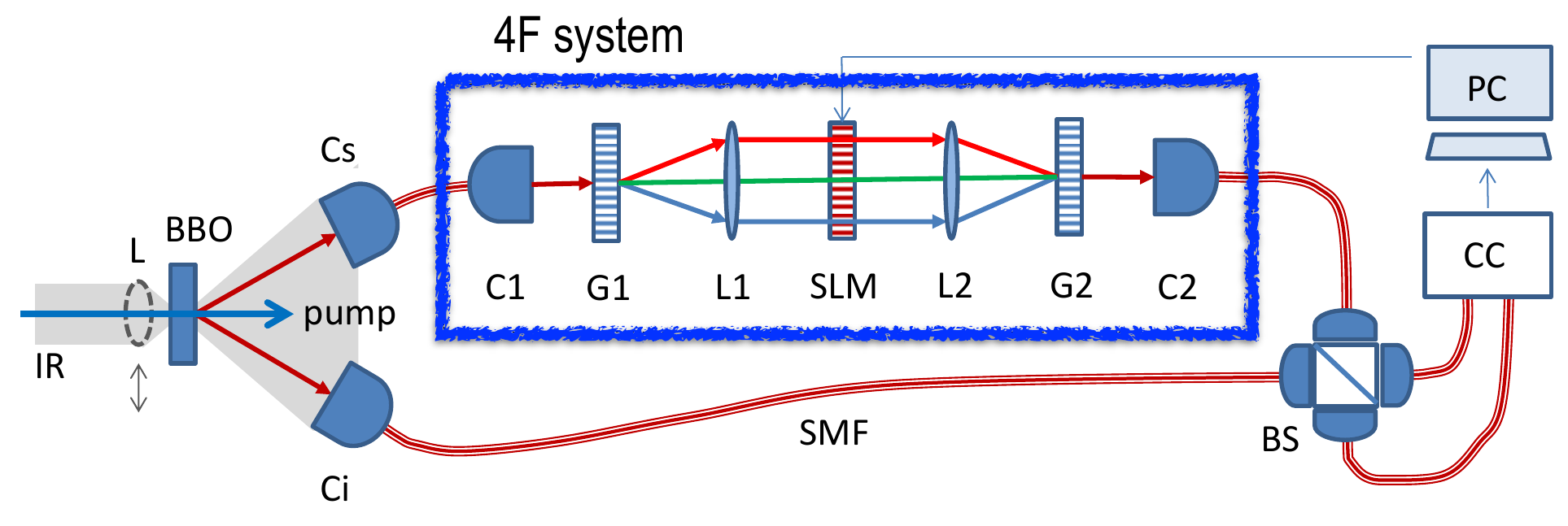}
\caption{\label{f:appar} Schematic diagram of our the two-photon interference
apparatus. From left to right, {the pump laser @ 405.5~nm (blue arrow)
enters the BBO crystal and is downconverted}. PDC photons are collected
with couplers (Cs, Ci).  The idler photon travels until the BS entirely
through the fiber (SMF). The signal photon travels also through the 4F
optical
system before entering again the fiber and reaching the BS. Two single
photon detectors, with a Time-to-Amplitude Converter/Single-Channel
Analyzer (TAC/SCA) system, detect the coincident counts (CC) at the two
BS outputs. {The shadowed region on the left, represents an IR
laser beam that a lens (L) focalizes in the BBO position in order to
perform an interference test}. The See the text for details.}
\end{figure}

In Fig.~\ref{f:appar} we show a schematic diagram of the experimental 
apparatus. The two frequency-entangled photon state is generated with a
pump laser @ $405.5$~nm by using a BBO crystal ($1$ mm thick). The
signal and idler photons are then collected by  two fiber couplers and
sent into single spatial mode and polarization 
preserving fibers (SMF). A $1$ mm thick crystal is enough to couple the 
maximum amounts of PDC radiation with our fiber couplers (Cs and Ci),
as the interaction length inside the crystal between the pump and the fiber
mode is slightly lower than $1$ mm \cite{tim9}.
The residual of the pump reaches a homemade
high-resolution spectrometer (not shown in the figure). The possibility
to stabilize the laser to different temperatures allows to keep
the same central wavelength while changing the current (and, thus,
the laser power) and also to perform two-photon
interference measurements with different pump laser widths (70~mA and
150~mA to have a narrower and broader pump respectively). When the idler photon
enters the coupler Ci, it travels entirely through the fiber towards
the BS. Conversely, the signal photon, after a short 
fiber, enters a 4F system \cite{wei2000}, i.e. propagates in the air, through
few optical devices (the couplers C1 and C2, the gratings G1 and G2 and lens L1 and L2) and a spatial light modulator (SLM).
At the end of the 4F system the signal photon is again coupled to a fiber
and finally reaches the BS, where it interferes with the idler. 
\par
The SLM is a 1D liquid crystal mask (640
pixels, 100~$\mu$m/pixel) and is placed on the Fourier plane between the two
lenses L1 and L2 of the 4F system (see Fig.~\ref{f:appar}). The SLM can be used
to introduce and control a phase function $\phi_{\text{M}}(\omegap,\omega)$
on the signal photon. The possibility to choose at will $\phi_{\text{M}}(\omegap,\omega)$
allows us to achieve several goals (even at the same time):
1) to introduce from second to high-order dispersion effects 
on the HOM dip; 2) to perform a delay-time scan of the coincidence 
counts rate, avoiding any mechanical stress on the apparatus 
(this method to perform  a delay scan is crucial 
for the stability of the  4F coupling efficiency); 3) to compensate 
any dispersion effects introduced by the fibers and the 4F system.
\par
As a matter of fact, also the presence of the 4F optical system on the signal path
introduces a phase function $\phi_{\text{4F}}(\omegap,\omega)$ and, in
particular, due to small  shifts of the grating positions with respect
to the optimal ones, it introduces {second  \cite{tim10} and} third
order dispersion{s} on its own.  \par
Finally, the signal and idler optical fibers
lead to the further phase functions $\phi_{\text{f,s}}(\omegap,\omega)$ and
$\phi_{\text{f,i}}(\omegap,\omega)$, respectively.
\par
Taking into account all the mentioned contributions, we have:
\begin{subequations}
\begin{align}
\phi_{\text{s}} (\omegap,\omega) &=
\phi_{\text{f,s}} (\omegap,\omega)+\phi_{\text{4F}} (\omegap,\omega)
+\phi_{\text{M}} (\omegap,\omega), \\[1ex]
\phi_{\text{i}} (\omegap,\omega) &=
\phi_{\text{f,i}} (\omegap,\omega),
\end{align}
\end{subequations}
and we can rewrite Eq.~(\ref{eq:Ptau}) as:
\begin{align}
\hspace{-4mm}P(\tau) =& \int\!\!\!\int\! d\omegap\, d\omega\, |A(\omegap)|^2 
\Big\{T^2 f(\omega)^2 + R^2 f(-\omega)^2 \nonumber\\
&-2\, R\, T\,\Re\text{e}\Big[f(\omega)\overline{f(-\omega)}
 \,\e^{\i \phi_{\text{tot}}(\omegap,\omega)} \Big]\Big\},\\
\approx& \int\!\!\!\int\! d\omegap\, d\omega\, |A(\omegap)|^2 
\Big\{T^2 f(\omega)^2 + R^2 f(-\omega)^2 \nonumber\\
&-2\, R\, T\,\Re\text{e}\Big[f(\omega)\overline{f(-\omega)}\nonumber\\
 &\hspace{1.5cm}\times\e^{\i 2\omega\tau+i\phi^{(2)}_{\text{tot}}(\omegap,\omega)+i\phi^{(3)}_{\text{tot}}(\omegap,\omega)}
 \Big]\Big\},
 \label{eq:ptausvolta}
\end{align}
where $\phi_{\text{tot}} (\omegap,\omega) =
\phi_{\text{s}} (\omegap,\omega)-\phi_{\text{s}} (\omegap,-\omega)+
\phi_{\text{i}} (\omegap,-\omega)-\phi_{\text{i}} (\omegap,\omega)$,
and the total second and third order dispersion terms are (note that, as mentioned,
the zero order terms cancel out):
\begin{align}
\phi^{(2)}_{\text{tot}}(\omegap,\omega)& = \beta^{(2)}_{\text{tot}}\, \omegap\, \omega, \\
\phi^{(3)}_{\text{\text{tot}}}(\omegap,\omega) & = \frac14\, \beta^{(3)}_{\text{tot}} 
\,\omegap^2\, \omega + 
\frac13\, \beta^{(3)}_{\text{tot}}\, \omega^3, \label{eq:phi3}
\end{align}
respectively, with:
\begin{equation}
\beta^{(k)}_{\text{tot}}= 
\beta_{\text{f}}^{(k)} + \beta_{\text{4F}}^{(k)} + 
\beta_{\text{M}}^{(k)} \qquad (k=2,3).
\end{equation}
According to our formalism, the quantities appearing at the RHS of the last equation
refer to the $k$-th order dispersion coefficients due to the fibers (f), to the 4F system (4F)
and to the SLM (M). It is worth noting that $\beta_{\text{f}}^{(k)} = d^{(k)}_{\text{f}} \Delta L$,
where $d^{(k)}_{\text{f}}$ is the $k$-th order dispersion of the fiber and $\Delta L$ is the length
difference between the signal and idler optical fiber.
\par
It is straightforward to see that in the presence of an almost monochromatic
pump, i.e., $\omegap \approx 0$, the only relevant dispersion effect comes from the
second term of the RHS in Eq.~(\ref{eq:phi3}), that is the second order dispersion does not affect 
the HOM dip shape.
\par
The presence of several optical components in the apparatus leads to an
overall loss in purity of the polarization quantum state of the photons. 
In order to assess the purity, we performed an interferometric 
measurement: after a strong filtering with neutral density 
filters, we focus a $811$ nm laser in the BBO position and couple 
the radiation of the resulting divergent beam with the fibers.
Then we obtain a Mach-Zehnder interferometer where the action of 
focusing plays the role of the first BS. Counts are then 
balanced by carefully positioning the focusing lens: this is crucial 
in order to have the same amount of direct counts on both paths, thus 
maximizing the interferogram visibility. 
\par
If  $\theta$ is the polarization mismatching between the two beams 
reaching the BS, the visibility of the
interferogram is given by $V_{\text{I}} = |\cos\theta |$.
It is now straightforward to show that the visibility of the HOM dip $V_{\text{H}}$ is
directly linked to the interferogram visibility by
\begin{equation}
V_{\text{H}}=\frac{R\,T\, V_\text{I}^2}{1-2\,R\,T-R\,T\,V_\text{I}^2}.
\end{equation}
In our experiment the measured visibility is $(95.6\pm 0.5)\%$, 
corresponding to a mismatching angle of 
$\theta = (17.2\pm 0.9)^{\circ}$ and a purity factor 
$p = \cos^2\theta \approx 91\%$.
Since, given $p$, the coincidence counts probability becomes:
\begin{equation}
\label{eq:purity}
P_{\text{tot}}(\tau) = P(\tau)\,p + \frac{1}{2}(1-p),
\end{equation}
then the maximum dip visibility  we can achieve with the measured
$T = 0.467$, $R = 0.533$ and a symmetric PDC spectrum,
i.e. $ f(\omega)=f(-\omega)$, turns out to 
be $(82\pm 2)\%$.
\par
\begin{figure}[tb]
\includegraphics[width=0.99\columnwidth]{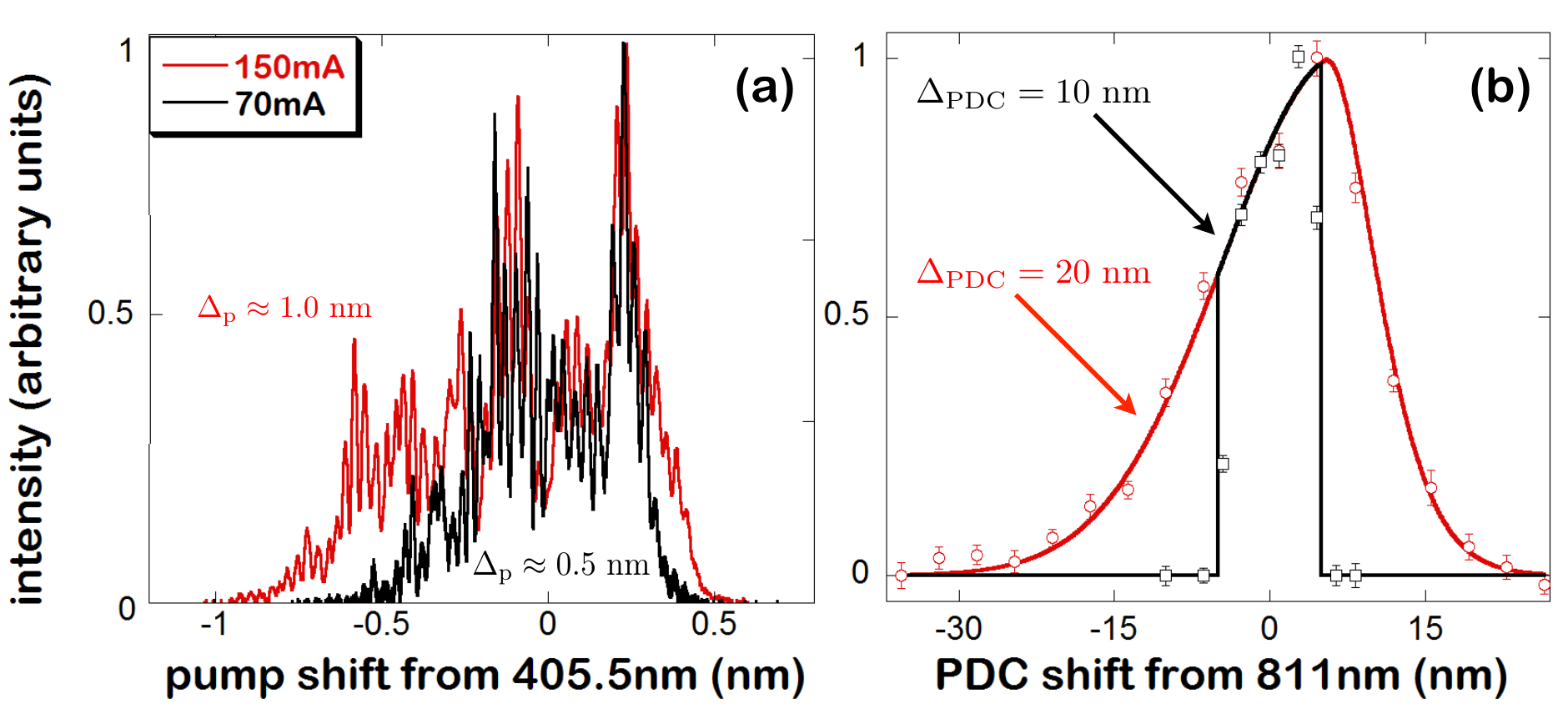}
\caption{\label{f:spectra} Pump and PDC spectra. 
Panel (a): measured pump spectrum shift about the
central wavelength of $405.5$ nm, with 70~mA (spectrum width $\Delta_{\text{p}} \approx 1.0$~nm)
and 150~mA (spectrum width $\Delta_{\text{p}} \approx 0.5$~nm) current. Panel 
(b): PDC spectrum data points and curve fit, with a slit on the
Fourier plane (spectrum width $\Delta_{\text{PDC}} = 10$~nm)
and without it (spectrum width $\Delta_{\text{PDC}} = 20$~nm).} \end{figure}
In Fig.~\ref{f:spectra} we show the pump and PDC spectra we used during
our measurements. On the pump side, in Fig.~\ref{f:spectra}(a) one can
see the difference in the pump spectrum width between two different current
settings: the narrower spectrum ($\Delta_{\text p}\approx 0.5$~nm) corresponds to
a current of $70$~mA, while a current of $150$~mA leads to a broader
spectrum ($\Delta_{\text p}\approx1$~nm). We measured the
pump spectrum with a homemade spectrometer a{s} follows. We collected the
residual of the pump light with a fiber and delivered it towards a collimating
lens. Then, a grating {($3600$~l/mm)} divides the collimated beam into
its spectral components. {These are then focused with a
$300$~mm lens onto the sensor of a CMOS camera placed on the Fourier
plane}.  \par
Figure~\ref{f:spectra}(b) shows the PDC spectrum. We can choose the
nominal spectrum ($\Delta_{\text{PDC}} = 20$~nm) or the ``cut'' spectrum
($\Delta_{\text{PDC}} = 10$~nm) with a slit placed on the Fourier plane
of the 4F system.  We measured PDC spectra with a $2$~mm slit on the
Fourier plane of the 4F system.  We calibrated the slit using a
graduated reference on the Fourier plane: we
found the $811$~nm component position for the slit, we measured a
dispersion coefficient of $3.62$~nm for a $2$~mm slit displacement and
finally, in order to measure the spectrum, for each slit position (and
therefore for each wavelength) we recorded coincidence counts from the
detectors.
\par
As mentioned above, one {usually} desire{s} to have symmetric PDC
spectra. However, Fig.~\ref{f:spectra}(b) shows that we used a non
symmetric signal photon spectrum. It is worth noting that our apparatus
allows us to center the PDC spectra on $811$ nm: this is possible by
transversally moving a mirror that delivers the pump inside the BBO
crystal. In this way we make the pump axis (that is the PDC cone axis)
be at the exactly same distance from both fiber couplers. However, we
found that the 4F system introduced an asymmetric cut on the PDC spectrum
tails, that is translated in an antisymmetric cut on the idler photon
spectrum. This makes the two photons distinguishable and the visibility
worsen. The spectrum we report in Fig.~\ref{f:spectra}(b) correspond to
the spectrum giving the highest visibility possible in our configuration:
the peaks are not superimposed at 811~nm but the tails are. Overall,
this makes the two photons less distinguishable than the situation where
the tails are not superimposed.
\section{Experimental results}\label{s:expr}
\begin{figure}[t]
\includegraphics[width=0.99\columnwidth]{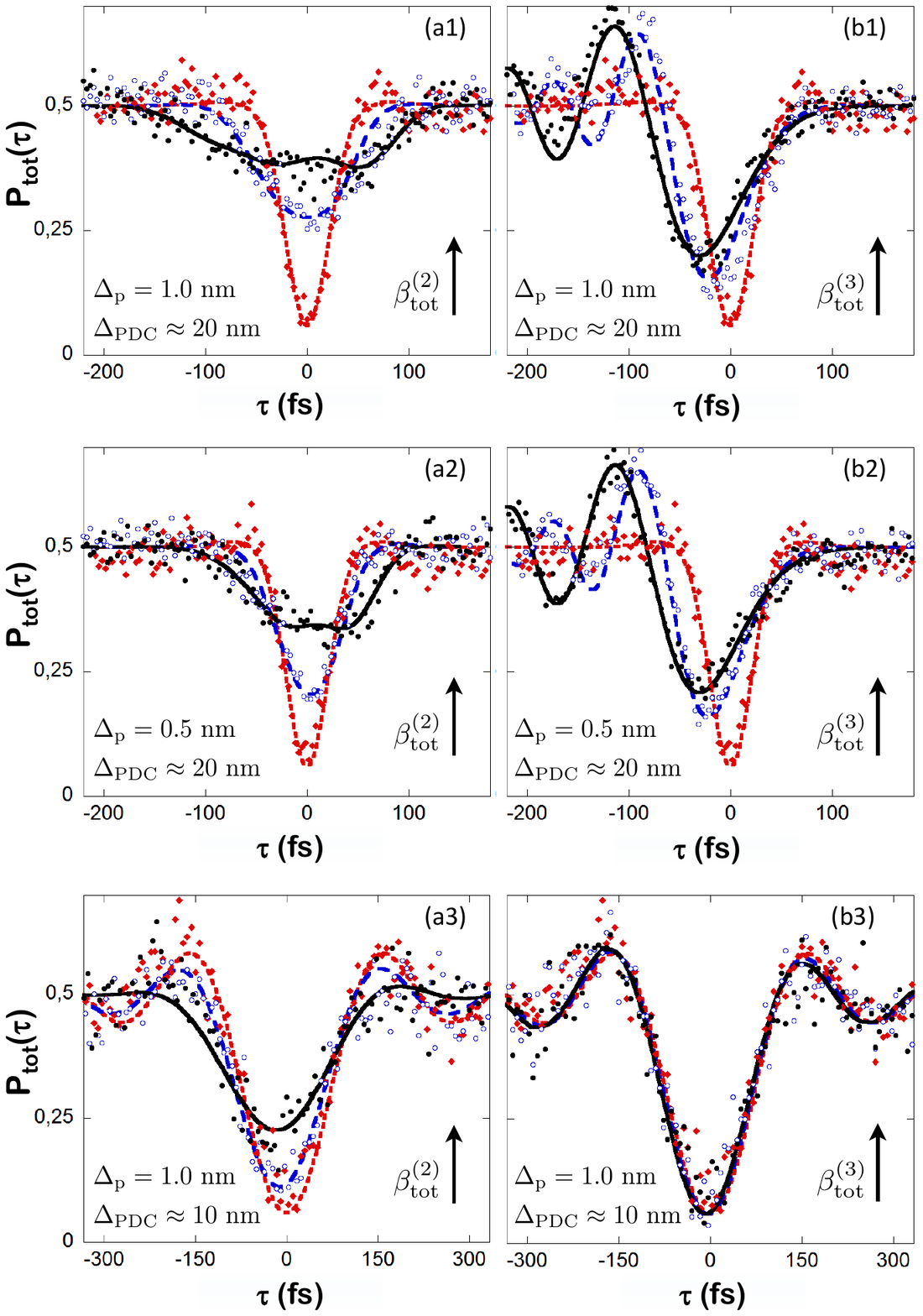}
\caption{\label{f:prof} The HOM dip in the presence of 
second and third order dispersion.  In panels (a1), (a2), and (a3) we
show the dip profiles for three different second order dispersion values,
from the lower to the higher dip we set $\beta^{(2)}_{\text{tot}} =0$~fs$^2$
(red dotted line and rhombus for theoretical prevision and
data points, respectively), $17.6\times10^3$~fs$^2$ (blue dashed line and
empty circles), $35.2\times10^3$~fs$^2$ (black solid line and filled circles).
In panels (b1), (b2), and (b3) we show the dip profiles for three different third order
dispersion values, from the lower to the higher dip we set
$\beta^{(3)}_{\text{tot}} = 0 $~fs$^3$ (red dotted line and rhombus),
$17.6\times10^4$~fs$^3$ (blue dashed line and empty circles),
$35.2\times10^4$~fs$^3$ (black solid line and filled circles).
The corresponding values of $\Delta_{\text p}$  and $\Delta_{\text{PDC}}$
of the pump and PDC spectrum widths are reported in the panels.
All solid lines are theoretical predictions obtained from Eq.~(\ref{eq:purity}).} 
\end{figure}

A slit placed under one of the fiber couplers is used to match signal
and idler paths. The SLM delay-time scan method allows us to match the
two paths with an error of less than $1$ $\mu$m. A range of several
picoseconds can be covered with an SLM delay-time scan. Therefore,
before measuring, signal and idler paths were always perfectly matched.
In Fig.~\ref{f:prof}, we report some measured dip profiles (points) together 
with the corresponding theoretical predictions (lines). 
It is worth noting that the predictions  are obtained from Eq.~(\ref{eq:purity})
and from the experimental spectra of Fig.~\ref{f:spectra} without
any fitting parameter. The experimental data are normalized using the
mean value at high delay time (i.e., outside the HOM dip). The time windows
for the counts acquisition for each point are $2$~s and $2.5$~s, respectively,
for the cases at $150$~mA and $70$~mA.
\par
Each line in Fig.~\ref{f:prof} corresponds to a different configuration.
On the left column, we show
three different profiles related to three different values of the second 
order dispersion coefficient; on the right column, we report
the profiles for three different values of the third order dispersion 
coefficient. 
In the top panels, (a1) and (b1), we have results for the 
broader pump spectrum ($1$~nm) and normal
PDC spectrum ($20$~nm) (from now on we refer to these values as to 
configuration S1).  We found a huge dependence on second and third  
order dispersion: the dip visibility decreases and its width increases, 
with increasing dispersion.
In the panels (a2) and (b2), we show results for the case of 
the narrower pump spectrum
($\Delta_{\text {p}} \approx 0.5$~nm) and normal PDC spectrum
($\Delta_{\text{PDC}} = 20$~nm, configuration S2).  In this case, the dependence of the
dip shape on the second order dispersion is significantly weaker, while
the dependency on the third order dispersion remains the same of the
previous case. This is in agreement with the fact that two-photon
interference is affected by second order dispersion only if the pump is
not monochromatic, while it is always affected by third order dispersion
(and all odd orders, including the first order).
\par
In the above cases, one may quantify the effects of the third order 
dispersion on the HOM dip by looking at the visibility. 
On the bottom line of Fig. \ref{f:prof}, panels (a3) and (b3), we have 
the case of the broader pump spectrum ($\Delta_{\text{p}} \approx 1.0$~nm) and the cut PDC 
spectrum ($\Delta_{\text{PDC}} = 10$~nm, configuration S3).  While we
still have a visible dependence on the second order dispersion, the
dependence on the third order dispersion disappears. The
width of the compensated dip, here, is much larger than in the cases
above. This comes directly from Eq.~\eqref{eq:ptausvolta}. A peculiar
behaviour can be seen here: symmetric oscillation appear on the dip,
caused by the clean cut of the PDC spectrum.

\section{Discussion}\label{s:discuss}
As we mentioned in the introduction, by a suitable choice of the
phase function we can simulate the high order dispersion effects of
a particular transmission channel.
In particualr, the values of the second and third order contribution we used in our
experiment are related to those obtained in current optical fibers.
For instance, a dispersion of  $30\times 10^3$~fs$^2$ is induced by a fused silica
fiber $\approx 850$~mm long ($d^{(2)}_{\text{f}}\approx 35$~fs$^2/\text{mm}$).
For a third order dispersion of $300\times
10^3$~fs$^3$, on the other hand, $\approx 11$~m of fiber are needed
($d^{(3)}_{\text{f}}\approx 30$~fs$^3/\text{mm}$) or, for example,
$35$~cm of ZnSe ($d^{(3)}_{\text{f}} = 870$ fs$^3/\text{mm}$).
It is worth noting that these lengths are quite small compared to those to be
used in telecommunications. For this reason the full comprehension of
these dispersion effects on the HOM dip becomes a fundamental a
prerequisite.
\begin{figure}[tb]
\includegraphics[width=0.99\columnwidth]{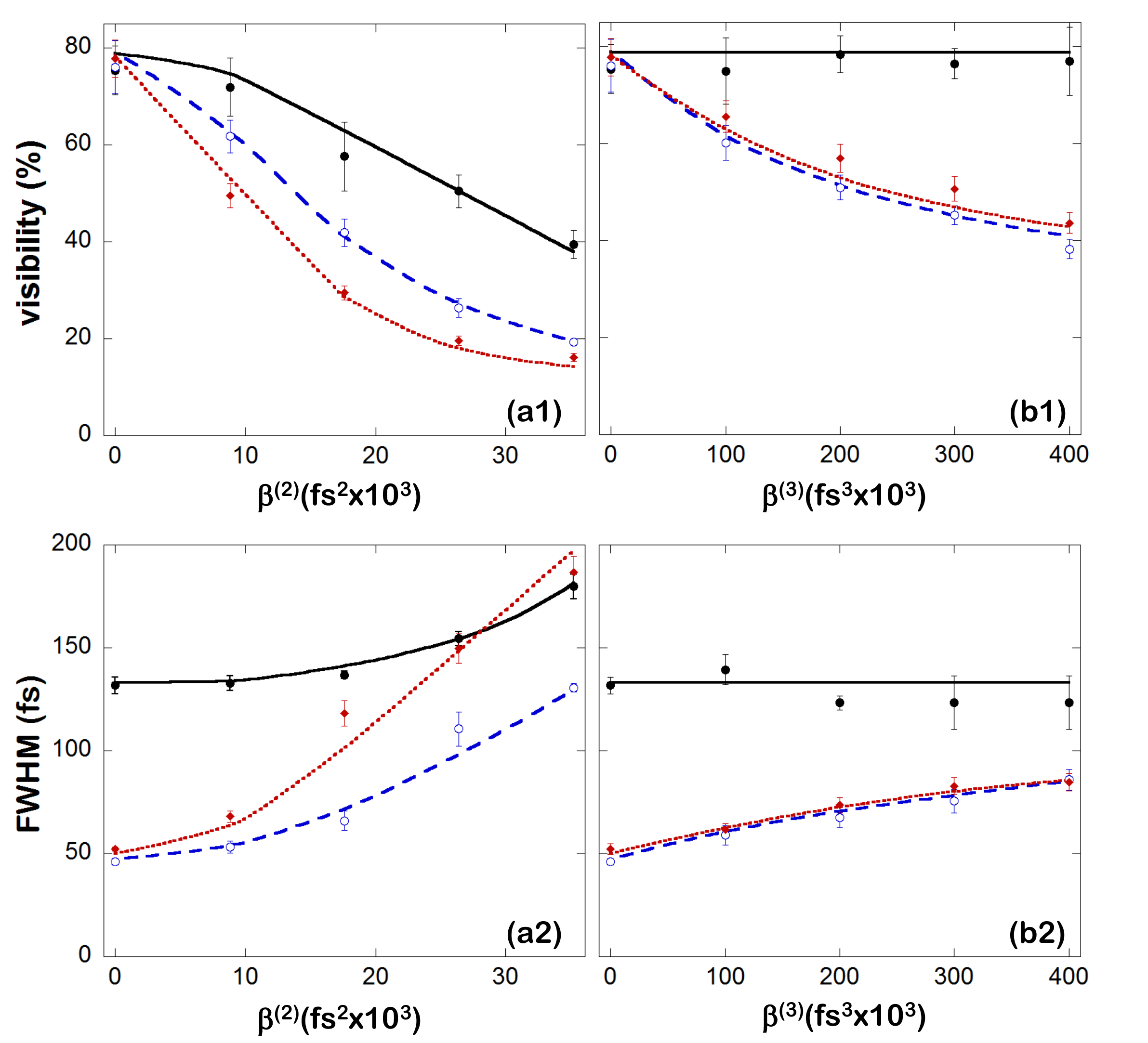}
\caption{\label{f:andamenti} 
Visibility and FWHM of the HOM dip as a function of
$\beta^{(2)}_{\text{tot}}$ and $\beta^{(3)}_{\text{tot}}$ for different
pump and PDC spectra:
PDC $\Delta_\text{PDC} = 10$~nm and $\Delta_\text{p} = 1.0$~nm (solid lines);
Dotted line: $\Delta_\text{PDC} = 20$~nm and $\Delta_\text{p} = 1.0$~nm (dotted);
$\Delta_\text{PDC} = 20$~nm and $\Delta_\text{p} = 0.5$~nm (dashed lines).}
\end{figure}
\par
In Fig.~\ref{f:andamenti} we summarize our main results: we can see the
behaviour of the dip visibility and full width at half maximum (FWHM) as
a function of second and third order dispersions. The visibility and the
FWHM are defined with respect to the normalized value and are obtained
by using suitable fitting functions.  Each plot reports the results for
the three different experimental configurations introduce above: dotted
red line for S1, dashed blue line for S2 and solid black line for S3. In
panel (a1), on the upper left, we see how visibility decreases as the
the second order dispersion increases. This worsening gets less sharp if
we cut down either the PDC spectrum (S3) or the pump spectrum (S2).  In
panel (a2),  on the lower left, we see how the dip FWHM increases (the
dip gets broader) when the second order dispersion coefficient
increases.  As already mentioned above, without dispersion the dip FWHM
in S3 is larger than in S1 and S2. In both S2 and S3, the FWHM is much
less sensitive to the second order dispersion  compared to the case of
broad pump and PDC spectra.
\par
In  Fig.~\ref{f:andamenti}(b1) and (b2), on the right, we studied again
the changes in the dip visibility and FWHM as a function of the third
order dispersion.  We can see how the cut of the PDC spectrum keeps
frozen the dip parameters, making them insensitive to any third order
dispersion coefficient. However, we can also clearly see that the
visibility and the FWHM do not depend on the pump spectrum.
\section{Conclusons}\label{s:out} 
In this paper we have designed and developed an 
experimental setup to simulate second and third 
order dispersion effects on the propagation of
PDC photons. By using our innovative apparatus, which is able to
introduce and manipulate a custom phase function with a SLM, we have
analyzed high-order dispersion effects in two-photon interference. In
particular, we have studied theoretically and experimentally, the
dependence of both the visibility and the width of the HOM dip on both
the pump spectrum and on the downconverted photon spectrum.  
\par 
Though we have mostly 
used our system to introduce high-order dispersion effects in the
propagation of the states, the same setup can be used to
\emph{compensate} such effect. Therefore, on the one hand, our results
clarify the role of the different dispersion coefficients and, on  the
other hand, pave the way for optimization procedures in communications
protocols and other applications in quantum
technology involving PDC photons and optical fibers.
\section*{Acknowledgments}
This work has been supported by UniMI 
through the UNIMI14 grant 15-6-3008000-609 and the H2020 Transition 
Grant 15-6-3008000-625, and by EU through the collaborative project 
QuProCS (Grant Agreement 641277). 

\end{document}